\documentclass[12pt]{article}
\usepackage{amssymb,amsmath,epsfig}
 \allowdisplaybreaks

\begin{document}
\title{\bf On the Stability of Pressure Isotropy Condition in Palatini $f(R)$ Gravity}
\author{Z. Tariq \thanks{zohatariq24@yahoo.com}, M. Z. Bhatti  \thanks{mzaeem.math@pu.edu.pk} and Z. Yousaf
\thanks{zeeshan.math@pu.edu.pk}\\
Department of Mathematics, University of the Punjab,\\
Quaid-e-Azam Campus, Lahore-54590, Pakistan.}
\date{}
\maketitle
\begin{abstract}
This manuscript copes with the issue of analyzing the conditions to check the stability of the pressure isotropy condition by taking into account a spherically symmetric dissipative astrophysical configuration with the Palatini $f(R)$ gravity theory. We work out a differential equation in terms of the Weyl scalar that has a crucial part in the analysis of evolution of considered system. Using this equation, we devise another stellar equation that characterize the evolution of anisotropic factor. Later, we assume an axially symmetric configuration and extended our analysis for that particular symmetry. It is worth-observing that the physical factors responsible for compelling an initially isotropic object to trigger pressure anisotropy incorporate energy density, dissipative flux and shear in fluid flow.
\end{abstract}

\section{Introduction}

From the recent astrophysical observations, researchers have found a confounding fact that the
universe is currently facing an accelerated expansion phase. After this revelation, scientists devoted their attention towards its possible elucidation. They found that may be our observable universe is homogeneously occupied by a fluid having negative pressure i.e., by dark energy (DE) counteracting the attractive force of gravitation of matter configurations. However, in Einstein's gravitation theory, i.e, General Relativity (GR), the notion of DE with visible matter cannot be carried out which makes it really difficult to formulate a theory that elucidates the cosmological evolution at all scales along with the description of origin of DE. From the up-to-date observations, almost $70$ percent of the total energy resources of our universe is DE and this is the reason that its feasible explication is vital. Another proposed explanation suggested that GR needs to be revised at large scale distances. Such revisions or modifications to GR are termed as modified theories of gravity (MTG). In these theories, the standard Einstein's-Hilbert action is revised so that the resulting modified field equations incorporate the terms that are crucial at low curvatures.

Over the past few decades, MTG have been put forward as an extensions to Einstein's gravitation theory. The prime inspiration behind using such gravitation theories is to explicate present-day astrophysical and cosmological data on dark matter (DM) and DE. Nojiri et al. \cite{1} analyzed some recent developments of MTG in cosmology with special emphasis on bouncing cosmology, inflation and accelerating era. By presenting the formalisms of $f(R),~f(T)$ and $f(G)$ theories of gravity, the authors also found certain astrophysical solutions and discussed their characteristics. Clifton et al. \cite{2} surveyed recent research in MTG along with their cosmological outcomes. They also reviewed the attempts of constructing parameterized post-Friedmannian technique which is helpful in constraining the deviations from GR. Capozziello et al. \cite{3} over-viewed the rudimentary principles of gravitational theories and discussed the geometrical interpretation by keeping in view the gauge theory. They focused, particularly, on the $f(R)$ and scalar-tensor theory using Palatini and metric approaches and analyzed the crucial role of torsion. Nojiri and Odintsov \cite{4} explored cosmological attributes and structure of a variety of modified theories and investigated relations and representations between them. They also reconstructed certain modified theories in detail and demonstrated that evolution of our universe may be re-examined for the theories under construction. They scrutinized big-rip and various other future singularities by taking MTG into account. Dombriz and Gomez \cite{5} focused on the analysis of certain attributes of modified gravity theories particularly the blackhole solutions by taking Friedmann-Lemaitre-Robertson-Walker spacetime. They further investigated the thermodynamical attributes of gravity theories of fourth order focusing on the global and local stability analysis of $f(R)$ models. Deep analysis of cosmological singularities is performed and entropy bounds are also authenticated.

The Palatini formalism suggests that in the variational principle, connection symbols and the metric tensor are geometric quantities with no dependence on each other. This kind of approach leads to the analysis of novel extensions to Einstein's general relativity in order to expound large scale astrophysical structures and dark energy characteristics. In particular, Palatini $f(R)$ gravitation theory ensures singularity-free second order Einstein's equations thus providing a gravitational substitute for dark energy. Olmo \cite{6} over-viewed the recent research on MTG in the background of Palatini formalism. The well-known problems including cosmic speedup problem and cauchy problem are deeply analyzed and the significance of using other models beyond $f(R)$ to comprehend DE and DM are discussed. Various attempts to solve the problem of cosmic speedup in $f(R)$ theories are also investigated. Barra\'{g}an et al. \cite{7} investigated the $f(R)$ cosmology using Palatini approach and analyzed certain restrictions sufficient for the existence of isotropic and homogeneous models. They demonstrated that a cosmic bounce can be used as a replacement for big-bang singularity for such models without the violation of energy conditions. It is also found that the proposed models can assist in avoiding the singularities during stellar collapse. Barra\'{g}an and Olmo \cite{8} investigated anisotropic and isotropic models in few MTG and took into account the bouncing non-singular solutions at early stages of the universe and found that $f(R)$ gravity models having bouncing isotropic solutions exhibit shear singularities if one considers pressure anisotropy. Yousaf et al. \cite{8a} performed stability analysis in order to check the homogeneity of the isotropic spherical distribution of matter. Wojnar \cite{9} constructed stellar Tolman-Oppenheimer-Volkoff equations for the modified theory utilizing the Palatini formalism keeping in view the spherically symmetric static geometry.

The analysis of stability and evolution of astrophysical systems after their formation has gained much significance in theoretical physics and it is considered as an intriguing phenomenon in MTG. The stellar configurations that are found to be stable against little fluctuations have significant physical attributes as compared to other models. Chan et al. \cite{10} scrutinized the part played by the pressure anisotropy and radiation on the instability of a stellar configuration. They concluded that for both Newtonian and post-Newtonian approximations, slight changes in anisotropy may effect the stability of the system significantly. Herrera et al. \cite{11} reattained the stellar evolution equation by taking into account a spherically symmetric, viscous, dissipative anisotropic fluid and analyzed the restrictions under which the shear-free condition remains stable. In particular, for the geodesic fluid, they found the possibility for the instability of shear-free condition. Herrera \cite{12} identified few elements causing inhomogeneity in energy density for the case of a self gravitating stellar system and analyzed the evolution of such elements by considering initially homogeneous matter distribution. For dissipative fluids, authors brought out the role of relaxational phenomena. Chan \cite{13} propounded a collapsing and radiating stellar configuration composed of isotropic fluid with shear stresses along with radial heat flow. The part played by various fluid variables like pressure, mass, luminosity and density etc. are examined and a comparison with previous relatable work in literature is established.

From the cosmological observations, it is clear that the fluid variables (shear stress, expansion scalar, vorticity tensor etc.) defining the kinematics of stellar configuration have substantial effects on its stability. Herrera et al. \cite{14} provided in-depth analysis of the physical meaning of scalar functions by taking into account a dissipative spherical fluid distribution with the inclusion of electromagnetic force along with the effects of cosmological constant. They investigated the part played by such scalar functions and also studied the physical outcomes. Letelier \cite{15} propounded a two-perfect-fluid model by taking into account anisotropic fluid and investigated the energy-momentum tensor associated with the model with focus on the stress tensor. A comparison between the Einstein's equations of the propounded model and a complex scalar field without mass is also carried out. Herrera and Santos \cite{16} extended the criteria of jeans instability to the stellar configurations having anisotropic stresses and found that instabilities may occur for the masses smaller than the corresponding jeans mass of local isotropic matter distribution. Herrera et al. \cite{17} analyzed instability for a spherically symmetric matter with anisotropic pressure collapsing adiabatically following restriction of zero expansion scalar by taking Newtonian and post-Newtonian regimes under consideration. They found that the instability range can be evaluated using pressure anisotropy in the fluid and radial depiction of energy density independent of stiffness.

A gravitational force responsible for holding together the constituent particles of a large relativistic stellar object is termed as self-gravitational force. In its absence, all the cosmological structures including huge galaxy clusters would fall into pieces. The investigation of self-gravitating astrophysical objects has enticed the focus of physicists. Prisco et al. \cite{18} scrutinized the cracking of self-gravitating compact relativistic stellar objects delineated through Raychaudhuri equation by using the time derivative of the expansion and by expressing the cracking occurrence criteria in integral form. Herrera \cite{19} investigated the cracking of a family of spheres resulting from the occurrence of tidal forces at various regions of the sphere and discussed certain applications to the astrophysical objects. Camm \cite{20} found that if we consider a system of equal mass stars, the Boltzmann equation takes the form of non-linear partial differential equation in three variables independent of each other. Three functions satisfying such equation are determined, two of which depict stellar systems in equilibrium which are found to be composed of self-gravitating or adiabatic gas. Herrera and Santos \cite{21} over-viewed the possible factors causing local anisotropy in self gravitating configurations and determined their effects on the system by considering both the general relativistic and Newtonian examples. The outcomes revealed that local anisotropy has a crucial part in studying the evolutionary stages of the propounded models. They also studied the significance of shear and the Weyl tensors and their connection with pressure anisotropy. Yousaf \cite{21a} and Bhatti \cite{21b} studied the stability of self-gravitating systems with the help of isotropic and anisotropic celestial objects.

Finding axially symmetric solutions for Einstein's stellar equations has been the center of attention for many researchers. Carot \cite{22} over-viewed the definition of axial symmetry and derived certain results explaining the geometry in the neighborhood of axis. By considering different coordinate systems, the expressions for line elements are also found. Mars and Senovilla \cite{23} investigated axially symmetric spacetimes having conformal symmetry and showed that in case of zero conformal symmetry, the conformal and axial killing vector commute. They found that the outcomes are independent of the Einstein's equations or any other fluid content. Hernandez \cite{24} obtained axially symmetric, static interior solutions to the Einstein's equations that match with the Weyl exterior solution. At matter surface, boundary conditions are employed which ensure that the coordinates having continuous metric coefficients along with their continuous derivatives exist. Herrera et al. \cite{25} gathered all the basic concepts necessary to investigate static axially symmetric sources. They wrote Einstein and conservation equations for the case of static fluid which is anisotropic in pressure and found certain exact analytical solutions, one of which exhibit an incompressible spheroid having isotropic pressure. Herrera and Jimenez \cite{26} studied axially symmetric spacetimes by taking Bondi metric and found certain structure scalars to determine the deviations from the spherical symmetry of a stellar object. They considered small departures from the spherical symmetry and specified the first order solution. Herrera et al. \cite{27} carried out an in-depth analysis on the collapse of axially symmetric matter by considering dissipative anisotropic fluid. A transport equation is also derived to study thermodynamical aspects of the system. Further, the structure scalars are procured and their role in the evolution of system is analyzed.

Taking into consideration the work of Herrera \cite{34} in GR, we have devised a framework to figure out the influence of $f(R)$ curvature corrections on the pressure isotropy condition using the Palatini approach. Our analysis is strictly parallel to the one presented in \cite{34}. The manuscript is structured as follows: In section \textbf{2}, the usual energy-momentum tensor and geometric terms for dark source along with all the necessary fluid variables are specified for the spherically symmetric case by taking into account the Palatini approach. Section \textbf{3} deals with a differential equation required for the comprehension of evolution for the Weyl scalar. In section \textbf{4}, we cope with the issue of determining the condition needed to make sure the validity of pressure isotropy condition throughout the evolutionary phases of the stellar configuration. In section \textbf{5}, we take into account the axially symmetric case and found the tetrad components of heat flux vector, four-acceleration and shear and vorticity tensor. Moreover, the components of heat transport equation are also determined. The obtained results are abridged in the discussion as section \textbf{6}. Lastly, section \textbf{7} incorporates some useful expressions used in our research.

\section{Modified Stellar Equations and Kinematical Variables}

The expression for Einstein-Hilbert action by keeping in view the $f(R)$ gravitational theory is
\begin{equation}\label{1b}
S_{f(R)}=\frac{1}{2\kappa}\int d^4x \sqrt{-g}f(R)+S_M,
\end{equation}
where $f(R)$ signifies a generic function for Ricci scalar, $\kappa$ symbolizes the coupling constant and $S_M$ indicates the matter action. Employing the technique of variation on Eq. (\ref{1b}) with $g_{\sigma\pi}$ and $\Gamma^\delta_{\sigma\pi}$, the following Palatini field equations are attained
\begin{align}\label{2b}
&\frac{1}{f_R}\left(\nabla_\sigma\nabla_\pi f_R-g_{\sigma\pi}\Box f_R\right)+\frac{\kappa T_{\sigma\pi}}{f_R}-\frac{1}{2}g_{\sigma\pi}
\left(R-\frac{f}{f_R}\right)+\frac{3}{2 f_R^2}\left[\frac{1}{2}g_{\sigma\pi}(\nabla f_R)^2\right.\\\nonumber &\left.-\nabla_\sigma f_R\nabla_\pi f_R\right]
=R_{\sigma\pi}-\frac{1}{2}g_{\sigma\pi}R,
\end{align}
with $\Box$ symbolizing the d' Alembert's operator. Another simple way of expressing Eq. (\ref{2b}) is
\begin{equation}\nonumber
G_{\sigma\pi}=\frac{\kappa}{f_R}( {T_{\sigma\pi}}+ \mathcal{T}_{\sigma\pi})\equiv
{T^{eff}_{\sigma\pi}},
\end{equation}
where $\mathcal{T}_{\sigma\pi}$ has the following value
\begin{align}\nonumber
\mathcal{T}_{\sigma\pi}&=\frac{1}{\kappa}\left(\nabla_\sigma\nabla_\pi f_R-g_{\sigma\pi}\Box f_R\right)+\frac{\kappa T_{\sigma\pi}}{f_R}-\frac{f_R}{2\kappa}g_{\sigma\pi}
\left(R-\frac{f}{f_R}\right)\\\nonumber&+\frac{3}{2\kappa f_R}\left[\frac{1}{2}g_{\sigma\pi}(\nabla f_R)^2 -\nabla_\sigma f_R\nabla_\pi f_R\right],
\end{align}
and the Einstein's tensor $G_{\sigma\pi}$ has the following form
\begin{equation}\nonumber
G_{\sigma\pi}=R_{\sigma\pi}-\frac{1}{2}g_{\sigma\pi}R.
\end{equation}
We assume a collapsing spherically symmetric fluid distribution locally anisotropic in pressure. Further, it is assumed that the fluid undergoes heat dissipation process. The metric can be written in the following form
\begin{equation}\nonumber
ds^2=-A^2(t,r)dt^2+B^2(t,r)dr^2+\hat{R}^2(t,r)(d\theta^2+sin^2\theta d\phi^2),
\end{equation}
where $A,~B$ and $\hat{R}$ are assumed to be only positive functions of radial and temporal coordinates. The energy-momentum tensor depicting the usual matter has the form
\begin{equation}\label{3b}
T_{\sigma\pi}=\mu V_\sigma V_\pi +Ph_{\sigma\pi}+\Pi_{\sigma\pi}+q(V_\sigma\chi_\pi+\chi_\sigma V_\pi),
\end{equation}
with
\begin{align}\nonumber
P=\frac{P_r+P_\bot}{3},\;\;\; h_{\sigma\pi}=g_{\sigma\pi}+V_\sigma V_\pi,\\\nonumber
\Pi_{\sigma\pi}=\Pi\left(\chi_\sigma\chi_\pi-\frac{h_{\sigma\pi}}{3}\right), \;\;\; \Pi=P_r -P_\bot.
\end{align}
By considering the observer and the fluid to be comoving, we can express four-velocity vector, heat flux vector along with a space-like four vector as
\begin{equation}\nonumber
V^\sigma=\frac{1}{A}\delta^\sigma_0, \;\;\; q^\sigma=\frac{q}{B}\delta^\sigma_1, \;\;\; \chi^\sigma=\frac{1}{B}\delta^\sigma_1,
\end{equation}
with $q$ being a function of $t$ and $r$.

\subsection{Kinematical Variables}

In order to comprehend the physical characteristics of the fluid, we work out certain important kinematical variables. In this regard, the components of four-acceleration defined as $a_\sigma=V_{\sigma;\pi}V^\pi$ along with its scalar $a^2=a^\sigma a_\sigma$ turn out to be
\begin{align}\nonumber
&a_0=\frac{\dot{f}_R}{2f_R}, \;\;\; a_1=\frac{A'}{A}+\frac{f_R'}{2f_R},\\\nonumber &a^2=\frac{A'^2}{A^2B^2}-\frac{{\dot{f}_R}^2}{4A^2f_R^2}+\frac{f_R'^2}{4B^2f_R^2}.
\end{align}
The expansion scalar defined by $\Theta=V^\sigma_{~;\sigma}$ and the components of shear tensor defined by $\tilde{\sigma}_{\sigma\pi}=V_{(\sigma;\pi)}+a_{(\sigma}V_{\pi)}-\frac{1}{3}\Theta h_{\sigma\pi}$ are acquired as
\begin{align}\nonumber
&\Theta=\frac{1}{A}\left[\frac{\dot{B}}{B}+\frac{2\dot{\hat{R}}}{\hat{R}}+\frac{2\dot{f}_R}{f_{R}}\right],\\\nonumber
&\tilde{\sigma}_{11}=\frac{2B\dot{B}}{3A}-\frac{B^2\dot{f}_R}{6Af_R}-\frac{2B^2\dot{\hat{R}}}{3A\hat{R}},\\\nonumber
&\tilde{\sigma}_{01}=\frac{Af_R'}{4f_R},\;\;\;\;\; \tilde{\sigma}_{22}=\frac{\hat{R}\dot{\hat{R}}}{3A}-\frac{\hat{R}^2\dot{f}_R}{6Af_R}-\frac{\dot{B}\hat{R}^2}{3AB},\\\nonumber
&\tilde{\sigma}_{33}=\tilde{\sigma}_{22}sin^2\theta.
\end{align}
The shear scalar in Palatini $f(R)$ gravity is
\begin{equation}\nonumber
\tilde{\sigma}^{\sigma\pi}\tilde{\sigma}_{\sigma\pi}=\frac{2}{3}\tilde{\sigma}^2+\frac{1}{4f_R^2}\left[\frac{\dot{f}_R^2}{3A^2}-\frac{f_R'^2}{2B^2}\right],
\end{equation}
where, we have
\begin{equation}\label{4b}
\tilde{\sigma}=\frac{1}{A}\left(\frac{\dot{B}}{B}-\frac{\dot{\hat{R}}}{\hat{R}}\right).
\end{equation}
Defining the velocity for the collapsing fluid $U$ as the rate of change of areal radius $R$ w.r.t. proper time, we can write $U=\dot{R}/A$.

\subsection{The Weyl tensor}

We can define the Weyl tensor $C^\rho_{\sigma\pi\mu}$ as a combination of its two parts, i.e., electric and magnetic. For the case of spherically symmetric object, the Weyl tensor can be defined only through its electric part as $E_{\sigma\pi}=C_{\sigma\mu\pi\nu}V^\mu V^\nu$ which has the following non-vanishing components
\begin{equation}\nonumber
E_{11}=\frac{2}{3}B^2\epsilon, \;\;\; E_{22}=-\frac{1}{3}\hat{R}^2\epsilon, \;\;\; E_{33}=E_{22}sin^2\theta,
\end{equation}
where $\epsilon$ signifying the Weyl scalar as
\begin{align}\nonumber
\epsilon&=\frac{1}{2A^2}\left[\frac{\ddot{\hat{R}}}{\hat{R}}-\frac{\ddot{B}}{B}-\frac{\dot{A}\dot{\hat{R}}}{A\hat{R}}
-\frac{\dot{\hat{R}}^2}{\hat{R}^2}+\frac{\dot{A}\dot{B}}{AB}
+\frac{\dot{B}\dot{\hat{R}}}{B\hat{R}}\right]+\frac{1}{2B^2}\left[\frac{A''}{A}-\frac{\hat{R}''}{\hat{R}}+\frac{B'\hat{R}'}{B\hat{R}}\right.\\\nonumber &\left.-\frac{A'B'}{AB}+\frac{\hat{R}'^2}{\hat{R}^2}-\frac{A'\hat{R}'}{A\hat{R}}\right].
\end{align}
Another way to express $E_{\sigma\pi}$ is
\begin{equation}\nonumber
E_{\sigma\pi}=\epsilon\left(\chi_\sigma \chi_\pi-\frac{1}{3}h_{\sigma\pi}\right).
\end{equation}
Utilizing the Palatini field equations, the following outcome for $\epsilon$ is attained
\begin{align}\label{5b}
\epsilon&=-\frac{4\pi\Pi}{f_R}+\frac{4\pi}{\hat{R}^3}\int_0^{\hat{R}}\left(\frac{\mu}{f_R}\right)'\hat{R}^3 d\hat{R}-\frac{12\pi}{R^3}\int_0^{\hat{R}}\left(\frac{qBU\hat{R}^2}{f_R}+\frac{\zeta_1\hat{R}'}{4\pi f_R}\right)d\hat{R}\\\nonumber &-(\zeta_2-\zeta_3).
\end{align}

\section{Stellar Evolution Equation for the Weyl Scalar $\epsilon$}

In this section, we acquire a differential equation already obtained in \cite{33} that interlinks the Weyl scalar and certain fluid variables to comprehend the evolution of astrophysical system. In our case, it reads
\begin{align}\label{6b}
\left[\epsilon+\frac{1}{f_R}(-4\pi\mu+4\pi\Pi+\zeta_4)\right]^.=\frac{3\dot{\hat{R}}}{\hat{R}}\left(\frac{4\pi}{f_R}(\mu+P_\bot)
-\epsilon\right)+\frac{12\pi qA\hat{R}'}{B\hat{R}f_R}+\zeta_5.
\end{align}
This differential equation exhibits that the basic fluid characteristics like pressure anisotropy and energy density affect the tidal force experienced by astrophysical object under consideration. We make use of this differential equation along with the conservation equation to acquire a stellar evolution equation for the anisotropic factor $\Pi$ which in turn provides the restrictions that guarantee time propagation of pressure isotropy.

\section{The Evolution of Isotropic Pressure Condition}

The well-known Tolman-Oppenheimer-Volkoff (TOV) equation in case of static spherically symmetric system can be acquired as
\begin{equation}\label{7b}
P_r'+(\mu+P_r)\frac{A'}{A}+2(P_r-P_\bot)\frac{\hat{R}'}{\hat{R}}+\phi_1=0.
\end{equation}
This is the well-known hydrostatic equilibrium equation that interconnects the pressure gradient $(P_r')$ with the gravitational force $(\mu+P_r)$ in the presence of pressure anisotropy $(P_r-P_\bot)$ along with the effects of modified gravity theory $(\phi_1)$. Now, presuming that the considered astrophysical system is compelled to abandon its equilibrium state, we imagine a picture of our system just after abandoning, i.e., at the time scale shorter than hydrostatic time, thermal adjustment and thermal relaxation time. So, at such time scale $q\approx U\approx\Theta\approx0 \Rightarrow \dot{\hat{R}}\approx \dot{B} \approx0$. Here, we assume that derivatives w.r.t. time of mentioned quantities are so small but not zero. Utilizing Palatini field equations along with the assumption that fluid is isotropic initially, we manipulate the anisotropic factor $\Pi$ at such time scale as
\begin{equation}\label{8b}
8\pi\Pi+f_R\left(\frac{\mathcal{T}_{11}}{B^2}-\frac{\mathcal{T}_{22}}{\hat{R}^2}\right)=\frac{1}{A}
\left(\frac{\ddot{B}}{B}-\frac{\ddot{\hat{R}}}{\hat{R}}\right)=\dot{\tilde{\sigma}}.
\end{equation}
It can be clearly observed that the system withdraws from the initial isotropic pressure condition within the considered time scale unless we consider the shear-free evolution of the fluid. Once a system withdraws its equilibrium state, it comes across two possible phases:
\begin{itemize}
  \item The fluid distribution becomes stable and retrieves a static state in the time scale of order of hydrostatic time.
  \item The fluid distribution becomes unstable and takes up a dynamic regime till it gets to its final equilibrium state.
\end{itemize}
In the first case, the anisotropic factor provided in Eq. (\ref{8b}) would not disappear in the newly attained equilibrium state and as a consequence, the system should exhibit pressure anisotropy whereas in the second case, we notice that the pressure isotropy condition is mandatory at any time scale, even if one considers the shear-free evolution of fluid distribution. By making use of Eq. (\ref{4b}) and second Bianchi identity mentioned in the Appendix, we may write the Eq. (\ref{6b}) as
\begin{align}\label{9b}
\frac{\partial}{\partial t}\left(\epsilon+\frac{4\pi\Pi}{f_R}\right)+\frac{\dot{\hat{R}}}{\hat{R}}\left(3\epsilon+\frac{4\pi\Pi}{f_R}\right)
&=-\frac{4\pi}{f_R}(\mu+P_r)A\sigma-\frac{4\pi Aq'}{Bf_R}-\frac{4\pi q}{Bf_R}\left(2A'-\frac{A\hat{R}'}{R}\right)+\vartheta_1,
\end{align}
which corresponds to Eq. (24) in \cite{34} with a difference that it incorporates the effects of curvature emendations denoted by $f_R$ and $\vartheta_1$.
We denote, for simplicity, the dissipative factor as $\Psi_{diss}$
\begin{equation}\nonumber
\Psi_{diss}=-\frac{4\pi}{Bf_R}\left(q(2A-\frac{A\hat{R}'}{\hat{R}})+Aq'\right).
\end{equation}
Equation (\ref{9b}) can be rewritten for the anisotropic factor as follows
\begin{align}\nonumber
\frac{\dot{\Pi}}{f_R}-\frac{\Pi \dot{f}_R}{f_R^2}+\frac{\dot{\hat{R}}\Pi}{\hat{R}f_R}=-\frac{1}{4\pi}
\left(\dot{\epsilon}+\frac{3\epsilon\dot{\hat{R}}}{\hat{R}}\right)-\frac{(\mu+P_r)A\sigma}{f_R}+\frac{\Psi_{diss}}{4\pi}+\frac{\vartheta_1}{4\pi}.
\end{align}
Again, this equation corresponds to Eq.(26) in \cite{34} but distinct from it in the sense that it incorporates an additional terms (second term on the left hand side and last term on the right hand side) emerging due to the curvature emendation. This when integrated along with the initial assumption $\Pi(t=0)=0$ produces the following outcome
\begin{align}\label{10b}
\Pi=-\frac{f_R}{4\pi \hat{R}}\int_0^t \hat{R}\left(\dot{\epsilon}+\frac{3\epsilon \dot{\hat{R}}}{\hat{R}}\right)d\tilde{t}-\frac{f_R}{\hat{R}}\int_0^t\frac{(\mu+P_r)A\sigma \hat{R}}{f_R}d\tilde{t}+\frac{f_R}{4\pi \hat{R}}\int_0^t\hat{R}(\Psi_{diss}+\vartheta_1)d\tilde{t},
\end{align}
which is parallel to Eq.(27) in \cite{34} with a major difference that here, the anisotropic factor is dependent also on the curvature emendations appearing due to alteration in the action integral.
From above equation, it is worth-noticing that mainly three factors are responsible for compelling the system to withdraw isotropic pressure condition. The first one being the effect of Weyl tensor, second being the shear force of the fluid flow and last term being the dissipative factor keeping in mind the consequences of Palatini $f(R)$ gravity denoted by $\vartheta_1$. We express the Weyl tensor in the above-mentioned equation in terms of fluid variables using Eq. ({\ref{5b}) and reattain Eq. (\ref{10b}) as
\begin{align}\nonumber
\Pi\dot{\hat{R}}&=\frac{f_R}{2\hat{R}^2}\frac{\partial}{\partial t}\int_0^{\hat{R}}\left(\frac{\mu}{f_R}\right)'\hat{R}^3d\hat{R}-\frac{3f_R}{2\hat{R}^2}\int_0^{\hat{R}}\left(\frac{qUB\hat{R}^2}{f_R}
+\frac{\zeta_1\hat{R}'}{4\pi f_R}\right)d\hat{R}\\\label{11b} &-\frac{\hat{R}f_R}{8\pi}\frac{\partial}{\partial t}(\zeta_2-\zeta_3)+\frac{(\mu+P_r)A\sigma \hat{R}}{2}-\frac{\hat{R}f_R\Psi_{diss}}{8\pi}-\frac{\vartheta_1\hat{R}f_R}{8\pi}.
\end{align}
From this equation, we can observe that the astrophysical configuration withdraws the isotropic pressure condition unless all the terms on the right-hand-side get canceled. Unlike Eq.(28) in \cite{34}, here the anisotropy factor $\Pi$ depends on $f(R)$ corrections terms (third and sixth term on the right hand side) which signifies that even if all the quantities vanish, the effects of gravity modification would still be present affecting the anisotropic factor.

\section{The Axially Symmetric Case}

The research done in preceding section can be diversified by taking into account an astrophysical configuration which is axially symmetric. We investigate the dynamics of such a stellar object just after it withdraws its equilibrium state. For such sources, the general metric is
\begin{equation}\nonumber
ds^2=-A^2(t,r,\theta)dt^2+2G(t,r,\theta)dtd\theta+B^2(t,r,\theta)dr^2+B^2(t,r,\theta)r^2d\theta^2+C^2(t,r,\theta)d\phi^2,
\end{equation}
with $A,~B,~C$ and $G$ being the positive functions of the said coordinates. As taken previously in Eq. (\ref{3b}), the fluid is dissipative and anisotropic which can be expressed by utilizing the same energy-momentum tensor. Choosing comoving coordinates again, we have
\begin{equation}\nonumber
V^\sigma=\left(\frac{1}{A},0,0,0\right), \;\;\;\;\; V_\sigma=\left(-A,0,\frac{G}{A},0\right).
\end{equation}
As defined in \cite{27} and \cite{34}, a canonical orthonormal tetrad $e_\sigma^{(\rho)}$ that adds three unitary space-like vectors to the four-velocity vector $e_\sigma^{(0)}=V_\sigma$ is
\begin{equation}\nonumber
e^{(1)}_\sigma=(0,B,0,0), \;\;\; e^{(2)}_\sigma=\left(0,0,\frac{\sqrt{B^2A^2r^2+G^2}}{A},0\right), \;\;\; e^{(3)}_\sigma=(0,0,0,C),
\end{equation}
with $\sigma=0,1,2,3$. The expression for anisotropic tensor defined previously in \cite{35} utilizing three scalar functions $\Pi_{(1)(1)},~ \Pi_{(2)(1)}$ and $\Pi_{(2)(2)}$ is
\begin{align}\nonumber
\Pi_{(2)(1)}&=e^\sigma_{(2)}e^\pi_{(1)}T^{(eff)}_{\sigma\pi},\\\nonumber
\Pi_{(2)(1)}&=\frac{1}{3}\left(2e^\sigma_{(1)}e^\pi_{(1)}-
e^\sigma_{(2)}e^\pi_{(2)}-e^\sigma_{(3)}e^\pi_{(3)}\right)T^{(eff)}_{\sigma\pi},\\\nonumber
\Pi_{(2)(2)}&=\frac{1}{3}\left(2e^\sigma_{(2)}e^\pi_{(2)}-
e^\sigma_{(3)}e^\pi_{(3)}-e^\sigma_{(1)}e^\pi_{(1)}\right)T^{(eff)}_{\sigma\pi}.
\end{align}
The heat flux vector exhibiting the heat dissipation is delineated by making use of two tetrad components stated as $q_{(1)}$ and $q_{(2)}$ as $q_\rho=q_{(1)}e^{(1)}_\rho+q_{(2)}e^{(2)}_\rho$. Using the coordinate form, we can express
\begin{equation}\nonumber
q^\rho=\left(\frac{q_{(2)}G}{A\sqrt{B^2A^2r^2+G^2}},\frac{q_{(1)}}{B},\frac{Aq_{(2)}}{\sqrt{B^2A^2r^2+G^2}},0\right),
\end{equation}
and
\begin{equation}\nonumber
q_\rho=\left(0,Bq_{(1)},\frac{\sqrt{B^2A^2r^2+G^2}q_{(2)}}{A},0\right).
\end{equation}
In a similar fashion, the components of four-acceleration may be written as combination of two scalar functions $a_{(1)}$ and $a_{(2)}$ as $a_\sigma=a_{(1)}e_\sigma^{(1)}+a_{(2)}e_\sigma^{(2)}$ with values of both scalar functions being equal to
\begin{align}\nonumber
a_{(1)}&=\frac{1}{B}\left(\frac{A'}{A}+\frac{f_R'}{2f_R}\right),\\\nonumber
a_{(2)}&=\frac{A}{\sqrt{B^2A^2r^2+G^2}}\left(\frac{\dot{G}^2}{A^2}-\frac{\dot{A}G}{A^3}+\frac{A_{,\theta}}{A}\right).
\end{align}
For the case of axially symmetric system, the expansion scalar turns out to be
\begin{equation}\nonumber
\Theta=\frac{1}{A}\left(\frac{2\dot{B}}{B}+\frac{\dot{C}}{C}\right)+\frac{G^2}{A(B^2A^2r^2+G^2)}
\left(-\frac{\dot{A}}{A}+\frac{\dot{G}}{G}-\frac{\dot{B}}{B}\right)+\frac{2\dot{f}_R}{Af_R}.
\end{equation}
The shear tensor can also be delineated by using two scalar functions $\sigma_{(1)(1)}$ and $\sigma_{(2)(2)}$ as
\begin{align}\nonumber
\tilde{\sigma}_{(1)(1)}&=\frac{1}{3A}\left(\frac{\dot{B}}{B}-\frac{\dot{C}}{C}\right)+
\frac{G^2}{3A(B^2A^2r^2+G^2)}\left(\frac{\dot{A}}{A}+\frac{\dot{B}}{B}-\frac{\dot{G}}{G}\right)-\frac{\dot{f}_R}{6Af_R},\\\nonumber
\tilde{\sigma}_{(2)(2)}&=\frac{1}{3A}\left(\frac{\dot{B}}{B}-\frac{\dot{C}}{C}\right)+
\frac{2G^2}{3A(B^2A^2r^2+G^2)}\left(\frac{\dot{A}}{A}+\frac{\dot{B}}{B}-\frac{\dot{G}}{G}\right)
-\frac{\dot{f}_R}{6Af_R}\\\nonumber &-\frac{G^2 f_{\dot{R}}}{2Af_R(B^2A^2r^2+G^2)}.
\end{align}
The scalar functions associated with vorticity tensor expressed as $\Omega_{\pi\rho}=\Omega_{(a)(b)}e^{(a)}_\pi e^{(b)}_\rho$ are
\begin{align}\nonumber
\Omega_{(1)(2)}&=-\frac{G}{2B\sqrt{B^2A^2r^2+G^2}}\left(\frac{G'}{G}-\frac{2A'}{A}-\frac{f_R'}{2f_R}\right)=-\Omega_{(2)(1)},\\\nonumber
\Omega_{(0)(1)}&=-\frac{f_R'}{4Bf_R}=-\Omega_{(1)(0)}.
\end{align}
while the value for $\Omega$ is evaluated as
\begin{equation}\label{12b}
\Omega=\frac{G}{2B\sqrt{B^2A^2r^2+G^2}}\left[\left(\frac{G'}{G}-\frac{2A'}{A}\right)^2-\frac{f_R'^2(B^2A^2r^2+G^2)}{4G^2f_R^2}\right]^{1/2}.
\end{equation}
Dissimilar to Eq. (46) in \cite{34}, the curvature emendation quantity (last term on right side of above expression) depicts that the scalar associated with the vorticity tensor is also affected if one considers modifications in the gravitational theory. In order to guarantee elementary flatness near the symmetry axis and at the center of astrophysical configuration, we need to impose the regularity conditions (\cite{22}, \cite{36}). Thus, as $r\approx0$, we have
\begin{equation}\label{13b}
\Omega=\sum_{i\geq1}\Omega^{(i)}(t,\theta)r^i,
\end{equation}
and in the vicinity of its center
\begin{equation}\label{14b}
G=\sum_{i\geq3}G^{(i)}(t,\theta)r^i.
\end{equation}

\subsection{The Transport Equation}

To comprehend the idea of thermodynamical evolution, we require a heat transport equation derived from the dissipative theory in \cite{28}-\cite{31}. The relaxation time denoted as $\bar{\tau}$ for the dissipative phenomenon is the most crucial parameter in such theories \cite{32}. It is delineated as the time consumed by the astrophysical object for getting back to its steady state after being removed from it. Thus, $\bar{\tau}$ has its own physical interpretation and if the time scale of the considered system gets smaller or equal to the order of $\bar{\tau}$, then neglecting this quantity would amount to the negligence of the whole system.
The heat transport equation reads
\begin{equation}\nonumber
\bar{\tau}h^\rho_\gamma q^\gamma_{;\pi}V^\pi+q^\rho=-\kappa h^{\rho\gamma}(T_{,\gamma}+Ta_\gamma)-\frac{1}{2}\kappa T^2\left(\frac{\bar{\tau} V^\sigma}{\kappa T^2}\right)_{;\sigma}q^\rho,
\end{equation}
where $\bar{\tau},~\kappa$ and $T$ signify relaxation time, thermal conductivity and temperature, respectively.
Applying the contraction on above equation with $e_\rho^{(2)}$, we attain
\begin{align}\label{15b}
&\frac{\bar{\tau}}{A}\left[\dot{q}_{(2)}+Aq_{(1)}\sqrt{\Omega^2+\frac{f_R'^2}{16B^2f_R^2}}
+\frac{\dot{f}_Rq_{(2)}}{2f_R}\right]+q_{(2)}=-\frac{\kappa}{A}\left[\frac{G\dot{T}+
A^2T_{,\theta}}{\sqrt{B^2A^2r^2+G^2}}\right.\\\nonumber &\left.+\frac{GT\dot{f}_R}{2f_R\sqrt{B^2A^2r^2+G^2}}
+ATa_{(2)}\right]-\frac{\kappa\bar{\tau}^2q_{(2)}}{2}\left(\frac{\bar{\tau} V^\sigma}{\kappa T^2}\right)_{;\sigma},
\end{align}
whereas applying contraction on same equation w.r.t. $e_\rho^{(1)}$, we acquire
\begin{align}\label{16b}
\frac{\bar{\tau}}{A}\left[\dot{q}_{(1)}-Aq_{(2)}\sqrt{\Omega^2+\frac{f_R'^2}{16B^2f_R^2}}
+\frac{\dot{f}_Rq_{(1)}}{2f_R}\right]+q_{(1)}&=-\frac{\kappa}{B}\left(T'+BTa_{(1)}\right)\\\nonumber&
-\frac{\kappa\bar{\tau}^2q_{(1)}}{2}\left(\frac{\bar{\tau} V^\sigma}{\kappa T^2}\right)_{;\sigma}.
\end{align}
Here, besides $\Omega$ (as appearing in Eqs. (50) and (51) in \cite{34}), we have the contributions from the modified gravity. Now, we discuss the characteristics of the system just after (i.e., the smallest time for which evolution can be noticed) its withdrawal from the equilibrium state. As provided earlier \cite{35}, the following observations are made.
\begin{itemize}
  \item When the time scale at which observer notices the system becomes smaller as compared to hydrostatic time scale, the values for the fluid variables $\Omega(G),~\Theta$ and the scalar functions $\tilde{\sigma}_{(1)(1)}$ and $\tilde{\sigma}_{(2)(2)}$ coincide with their values in equilibrium state. So, these values may be disregarded but not their derivatives w.r.t. time that are supposed to be of very small order.
  \item Heat flux vector must be disregarded exclusive of its time derivative.
  \item From both the above-mentioned observations, it is clear that the first order derivatives w.r.t. temporal component of $A,~B$ and $C$ can also be disregarded.
\end{itemize}
Following the above criteria, we obtain for four acceleration
\begin{equation}\nonumber
a_{(1)}=\frac{A'}{AB}+\frac{f_R'}{2Bf_R}, \;\;\; a_{(2)}=\frac{1}{Br}\left(\frac{A_{,\theta}}{A}+\frac{\dot{G}}{A^2}\right).
\end{equation}
The other kinematical quantities take the following form
\begin{align}\nonumber
\dot{\Theta}&=\frac{1}{A}\left(\frac{2\ddot{B}}{B}+\frac{\ddot{C}}{C}+\frac{2\ddot{f}_R}{f_R}-\frac{2\dot{f}_R^2}{f_R^2}\right),\\\nonumber
\dot{\tilde{\sigma}}_{(1)(1)}&=\dot{\tilde{\sigma}}_{(2)(2)}=\dot{\tilde{\sigma}}=\frac{1}{3A}\left(\frac{\ddot{B}}{B}-
\frac{\ddot{C}}{C}-\frac{\ddot{f}_R}{2f_R}+\frac{\dot{f}_R^2}{2f_R^2}\right),\\\nonumber
\dot{\Omega}&=\frac{1}{AB^2r}\left(\frac{\dot{G}^2}{2}-\frac{A'\dot{G}}{A}-\frac{f_r'\dot{f}_R'B^2A^2r^2}{8G'f_R^2}
+\frac{\dot{f}_Rf_R'^2B^2r^2A^2}{8G'f_R^3}\right).
\end{align}
When the astrophysical system reaches its state of thermal equilibrium, the heat flux vector becomes zero and the Tolman conditions for such thermal equilibrium $(TA)'=(TA)_{,\theta}=0$ are fulfilled.
We manipulate Eq. (\ref{15b}) and (\ref{16b}) just after the withdrawal of the system from equilibrium and obtain following outcomes respectively.
\begin{align}\label{17b}
\dot{q}_{(1)}&=\frac{-\kappa TAf_R'}{2B\bar{\tau}f_R},\\\nonumber
\bar{\tau}\dot{q}_{(2)}&=-\frac{\kappa AT_{,\theta}}{Br}-\kappa ATa_{(2)}.
\end{align}
Here, utilizing the Tolman condition for thermal equilibrium, we acquire
\begin{equation}\label{18b}
\bar{\tau}\dot{q}_{(2)}=-\frac{\kappa T\dot{G}}{ABr}.
\end{equation}
Next, we make use of the above observations and evaluate non-vanishing Einstein's tensor components for axially symmetric case just after withdrawal of system from its equilibrium state. At such time scale, the Einstein's tensor incorp components orates following three kinds of terms.
\begin{itemize}
  \item The factors with first order time derivatives of the metric variables that we have taken to be equal to zero.
  \item The terms representing the equilibrium state of the system i.e., neither containing $G$ nor the first time derivatives of metric variables.
  \item The terms containing the first order time derivative of $G$ or second order derivatives of metric variables $A,~B$ and $C$ w.r.t. time which can not be disregarded.
\end{itemize}
Thus, from the Palatini field equations, we attain the following outcomes
\begin{align}\label{19b}
8\pi\left(\frac{\mu}{f_R}+\mathcal{T}_{00}\right)&=8\pi\left(\frac{\mu}{f_R}+\mathcal{T}_{00}\right)_{(eq)},\\\nonumber
8\pi\left(\frac{P}{f_R}+\mathcal{T}_{11}\right)&=8\pi\left(\frac{P}{f_R}+\mathcal{T}_{11}\right)_{(eq)}-\frac{2\dot{\Theta}}{3A}
+\frac{4\ddot{f}_R}{3A^2f_R}-\frac{4\dot{f}_R^2}{3A^2f_R^2}\\\label{20b}&+\frac{2}{3A^2B^2r^2}
\left(\dot{G}_{,\theta}+\frac{\dot{G}C_{,\theta}}{C}\right),\\\nonumber
8\pi\left(\frac{\Pi_{(1)(1)}}{f_R}+\mathcal{T}_{11}\right)&=8\pi\left(\frac{\Pi_{(1)(1)}}{f_R}+
\mathcal{T}_{11}\right)_{(eq)}+\frac{\dot{\tilde{\sigma}}}{A}+\frac{\ddot{f}_R}{6A^2f_R}-\frac{\dot{f}_R^2}{6A^2f_R^2}\\\label{21b}&+\frac{1}{3A^2B^2r^2}
\left(\dot{G}_{,\theta}-\frac{3\dot{G}B_{,\theta}}{B}+\frac{\dot{G}C_{,\theta}}{C}\right),\\\nonumber
8\pi\left(\frac{\Pi_{(2)(2)}}{f_R}+\mathcal{T}_{22}\right)&=8\pi\left(\frac{\Pi_{(2)(2)}}{f_R}+
\mathcal{T}_{22}\right)_{(eq)}+\frac{\dot{\tilde{\sigma}}}{A}+\frac{\ddot{f}_R}{6A^2f_R}-\frac{\dot{f}_R^2}{6A^2f_R^2}\\\label{22b}&+\frac{1}{3A^2B^2r^2}
\left(-2\dot{G}_{,\theta}+\frac{3\dot{G}B_{,\theta}}{B}+\frac{\dot{G}C_{,\theta}}{C}\right),\\\nonumber
8\pi\left(\frac{\Pi_{(2)(1)}}{f_R}+\mathcal{T}_{21}\right)&=8\pi\left(\frac{\Pi_{(2)(1)}}{f_R}+
\mathcal{T}_{21}\right)_{(eq)}-\frac{\dot{\Omega}}{A}-\frac{f_R'\dot{f}_R'}{8A^2B^2G'rf_R}\\\label{23b} &+\frac{\dot{f}_Rf_R'^2}{16A^2B^2G'rf_R^2}+\frac{\dot{G}}{A^2B^2r}
\left[\frac{(Br)'}{Br}-\frac{A'}{A}\right].
\end{align}
The symbol $(eq)$ written in the subscript of certain quantities are the values of those particular quantities in equilibrium. Further, we suppose that the pressure in our system is isotropic initially i.e., $\Pi_{(1)(1)(eq)}=\Pi_{(2)(2)(eq)}$ and $\Pi_{(2)(1)(eq)}=0$. In order to determine whether these restrictions propagate in time or not, we presume that out of the equilibrium state of the system, we have $\Pi_{(1)(1)}=\Pi_{(2)(2)}$. From Eqs. (\ref{21b}) and ({\ref{22b}}), we acquire
\begin{equation}\nonumber
\dot{G}=B^2\tilde{f}(t,r).
\end{equation}
Here, $\tilde{f}$ corresponds to the fluid news function as provided in \cite{35}. Making an appropriate choice for this function, we observe that the regularity conditions are satisfied and no conflict with any other stellar equation occurs. Keeping this in mind, we may suppose that even after abandoning the equilibrium state, the system fulfils the criteria of $\Pi_{(1)(1)}=\Pi_{(2)(2)}$ but this is not true for the case of $\Pi_{(2)(1)}$. If we consider $\Pi_{(2)(1)}=0$, then from Eq. (\ref{23b}), we procure
\begin{equation}\nonumber
\frac{\dot{\Omega}}{A}=\frac{\dot{G}}{A^2B^2r}\left[\frac{(Br)'}{Br}-\frac{A'}{A}\right]-\frac{f_R'\dot{f}_R'}
{8A^2B^2G'rf_R}+\frac{\dot{f}_Rf_R'^2}{16A^2B^2G'rf_R^2}.
\end{equation}
This outcome is different from Eq.(65) in \cite{34} because of the fact that the time derivative scalar function associated with the vorticity tensor $\Omega$ is directly dependent on the $f(R)$ terms. Using the above equation along with Eq.(\ref{12b}), we acquire
\begin{equation}\label{24b}
\dot{G}=B^2r^2\tilde{g}(t,\theta),
\end{equation}
with $\tilde{g}$ being an arbitrary function. Notice that Eq.({\ref{24b}}) does not satisfy the regularity condition in the vicinity of center. Thus, for the considered time scale, we should have $\Pi_{(2)(1)}\neq0$. Keeping this in view, we attain
\begin{equation}\nonumber
8\pi\left(\frac{\Pi_{(2)(1)}}{f_R}+\mathcal{T}_{21}\right)=\frac{\tilde{f}}{2A^2r}\left(ln \frac{r^2}{\tilde{f}}\right)'.
\end{equation}
This depicts that after the withdrawal of the astrophysical system from the equilibrium state, $\Pi_{(1)(1)}=\Pi_{(2)(2)}$ is satisfied but the case is not the same for $\Pi_{(2)(1)}$. As the system withdraws its equilibrium state, the function $\tilde{f}$ administers its evolution. So, $\tilde{f}$ cannot be zero which in turn exhibits that $\Pi_{(2)(1)}$ should also be non-zero (along with the contribution of dark source term $\mathcal{T}_{21}$ absent in Eq.(67) in \cite{34}) even if it vanishes initially. It is worth-noticing from Eqs.(\ref{18b}) and (\ref{22b}) that the function $\tilde{f}$ prompts the dissipation process. For the case of spherical symmetry, such physical quantities are responsible for bringing on anisotropy in pressure when the system leaves its initial state with pressure isotropy in it.

If the astrophysical system comes back to static state in the time of order of hydrostatic time, it keeps the non-zero value for $\Pi_{(2)(1)}$ which it attained after withdrawing equilibrium state. This exhibits that in the newly attained static state, the distribution would necessarily be anisotropic.

\section{Discussion}

The gravity theories based on Palatini variation are superior to others in the sense that they provide the right Newtonian limit and pass all the solar system tests accurately. The Palatini theories explicate the effective spacetime geometry with the quantum structure which in turn assist in comprehension of recent accelerated expansion phase of our universe. Such theories do not incorporate instabilities and are a source of non-linear second order evolution equations based on a basic notion that there is no prior connection between metric and affine connection $\Gamma^\delta_{\sigma\pi}$. The affine connection includes the effects of dark source terms and for this reason, the kinematical variables e.g., expansion scalar and shear tensor etc. exhibit the contributions from the dark source ingredients.

It is a widely known fact that the physical processes responsible for inducing pressure anisotropy in any system are found in compact astrophysical configurations. Also, much research has been carried out regarding a relativistic fluid with isotropic pressure. The matter of our concern is to find the restrictions under which initially isotropic fluid distribution remains as it is while undergoing the phases of evolution by bearing in mind the Palatini $f(R)$ gravity theory.

When a fluid configuration withdraws the pressure isotropy condition, it becomes  stable and returns to its equilibrium state after being removed from it. When the system attains this new equilibrium state, it would necessarily be anisotropic. The kinematical factors e.g., dissipative flux, inhomogeneous energy density, shear tensor and the integral defined in first term of Eq. (\ref{11b}) along with the dark source ingredients induce the occurrence of anisotropy in pressure. From this equation, it is clear that the isotropic fluid distribution must be shear-free, non-dissipative having homogeneous energy density in order to remain isotropic throughout the evolution process. We can observe from this equation that another possibility for the fluid to remain isotropic is that the terms specified on the right-hand-side get canceled. As far as the stability of the isotropic pressure condition is concerned, Eq. (\ref{10b}) exhibit that vanishing shear forces, absence of dissipation, conformal flatness (i.e., zero Weyl tensor) and contribution of the dark source quantities are the factors that play a crucial role. It is worth-noticing from Eq. (\ref{11b}) that even if we suppose pressure isotropy, homogeneous energy density and shear-free fluid, the dissipation factor would improve the withdrawal from the pressure isotropy condition. In a nutshell, it is found that non-dissipation, dark source ingredients, conformal flatness and shear-free restrictions throughout the evolutionary phase are enough to guarantee that considered matter evolves while maintaining the pressure isotropy condition at each time.

The same analysis is carried out for the case of axially symmetric configurations and it came out that the evolution of an initially isotropic fluid, at the considered time scale, eventually leads to anisotropic fluid. We observe that even if the functions $\Pi_{(1)(1)}$ and $\Pi_{(2)(2)}$ are equal after withdrawal from equilibrium, the third one i.e., $\Pi_{(2)(1)}$ should necessarily be non-vanishing. In such a non-spherical collapse, in addition to the intense dissipation process, pressure anisotropy along with the contributions of modified gravity terms play a significant role as the stellar object reaches its final stages of evolution. We conclude that a fluid configuration  which is isotropic at its initial stage would generate pressure anisotropy while experiencing the stages of evolution. The magnitude of the attained anisotropy would, in particular, depend on the data of the stellar configuration.

\section{Appendix}

The components of Bianchi identities are
\begin{align}\nonumber
T^{(eff)\sigma\pi}_{;\pi}\chi_\sigma&=\frac{1}{A}\left[\dot{q}+2q\left(\frac{\dot{B}}{B}
+\frac{\dot{\hat{R}}}{\hat{R}}\right)\right]+\frac{1}{B}\left[P_r'+(\mu+P_\bot)\frac{A'}{A}+2(P_r-P_\bot)\frac{\hat{R}'}{\hat{R}}\right]+\psi_1,\\\nonumber
T^{(eff)\sigma\pi}_{;\pi}V_\sigma&=-\frac{1}{A}\left[\dot{\mu}+(\mu+P_r)\frac{\dot{B}}{B}
+2(\mu+P_\bot)\frac{\dot{\hat{R}}}{\hat{R}}\right]-\frac{1}{B}\left[q'+2q\frac{(A\hat{R})'}{A\hat{R}}\right]+\psi_2.
\end{align}
The values for the variables used in the expressions for spherically symmetric case are
\begin{align}\nonumber
\zeta_1&=-\frac{-9\hat{R}^2\dot{f}_R^2}{4A^2f_R}-\frac{\hat{R}^2f_R'^2}{8B^2f_R}+\frac{\hat{R}^2f_R''}{2B^2}
-\frac{\hat{R}^2\dot{B}\dot{f}_R}{2A^2B}-\frac{B'\hat{R}^2f_R'}{2f_R}-\frac{\hat{R}\dot{\hat{R}}\dot{f}_R}{A^2}+\frac{\hat{R}\hat{R}'f_R'}{B^2}\\\nonumber
&-\frac{\hat{R}^2}{4}(f-Rf_R)-\frac{\dot{\hat{R}}\hat{R}^2\dot{f}_R'}{2A^2\hat{R}'}+\frac{\hat{R}\hat{R}^2A'\dot{f}_R}{2A^3\hat{R}'}
+\frac{\dot{B}\dot{\hat{R}}\hat{R}^2f_R'}{2A^2B\hat{R}'}+\frac{5\dot{f}_Rf_R'\dot{\hat{R}}\hat{R}^2}{4A^2\hat{R}'f_R},\\\nonumber
\zeta_2&=\frac{9\dot{f}_R^2}{4A^2f_R}+\frac{f_R'^2}{8B^2f_R}-\frac{f_R''}{2B^2}+
\frac{\dot{B}\dot{f}_R}{2A^2B}+\frac{B'f_R'}{2f_R}+\frac{\dot{\hat{R}}\dot{f}_R}{A^2\hat{R}}-\frac{\hat{R}'f_R'}{B^2\hat{R}}+\frac{1}{4}(f-Rf_R),\\\nonumber
\zeta_3&=\frac{5f_R'^2}{4B^2f_R}-\frac{\dot{A}\dot{f}_R}{A^3}-\frac{\dot{\hat{R}}\dot{f}_R}{2A^2\hat{R}}+\frac{\hat{R}'f_R'}{2B^2\hat{R}}
+\frac{f_R''}{2B^2}-\frac{\dot{B}\dot{f}_R}{2A^2B}-\frac{B'f_R'}{2B^3},\\\nonumber
\zeta_4&=\zeta_2-\zeta_3,\\\nonumber\zeta_5&=\frac{3}{2B^2\hat{R}f_R}(\dot{\hat{R}}\tau_2-\hat{R}'\tau_1)-\frac{3\dot{\hat{R}}}{\hat{R}}(\zeta_2-\zeta_3).
\end{align}
The expressions for the dark source ingredients are
\begin{align}\nonumber
\tau_1&=\dot{f}_R'-\frac{A'\dot{f}_R}{A}-\frac{\dot{B}f_R'}{B}-\frac{5\dot{f}_Rf_R'}{2f_R},\\\nonumber
\tau_2&=-\frac{B^2\dot{f}_R^2}{4A^2f_R}-\frac{9f_R'^2}{4f_R}+\frac{B^2\ddot{f}_R}{A^2}+\frac{\dot{A}B^2\dot{f}_R}{A^3}
-\frac{A'f_R'}{A}+\frac{2B^2\dot{\hat{R}}\dot{f}_R}{A^2\hat{R}}-\frac{2\hat{R}'f_R'}{R}\\\nonumber&+\frac{B^2}{2}(f-Rf_R).
\end{align}
Also,
\begin{align}\nonumber
\phi_1&=\frac{1}{\kappa}\left[\frac{A'f_R''}{AB^4}-\frac{4A'f_R'^2}{AB^4f_R}-\frac{3A'B'f_R'}{AB^5}
+\frac{3f_R'f_R''}{B^4f_R}-\frac{6f_R'^3}{B^4f_R^2}-\frac{15B'f_R'^2}{B^5f_R}+\frac{f_R'(f-Rf_R)}{2B^5f_R}\right.\\\nonumber &\left.
-\frac{A'^2f_R'}{A^2B^4}+(T^{11})'-\frac{6B'\hat{R}'f_R'}{B^5\hat{R}}+\frac{B'(f-Rf_R)}{B^3}-\frac{8\hat{R}'f_R^2}{B^4\hat{R}f_R}
-\frac{2\hat{R}'^2f_R'}{B^4\hat{R}^2}+\frac{2B'f_R''}{B^4\hat{R}}\right],\\\nonumber
\psi_1&=-A(\mathcal{T}^{00})^.-A(\mathcal{T}^{01})'+\mathcal{T}^{00}\left(-2\dot{A}-\frac{A\dot{B}}{B}-\frac{A\dot{\hat{R}}}{\hat{R}}-\frac{2A\dot{f}_R}{f_R}\right)\\\nonumber&
+\mathcal{T}^{10}\left(3\dot{B}+\frac{\dot{A}B}{A}+\frac{\dot{\hat{R}}B}{\hat{R}}+\frac{5B\dot{f}_R}{2f_R}\right)+\mathcal{T}^{11}\left(2B'+\frac{A'B}{A}
+\frac{\hat{R}'B}{\hat{R}}+\frac{2Bf_R'}{f_R}\right)\\\nonumber&+\mathcal{T}^{22}\left(\frac{-2\hat{R}\hat{R}'}{B}-\frac{\hat{R}^2f_R'}{Bf_R}\right),\\\nonumber
\vartheta_1&=\zeta_5-(\zeta_4)^.+\frac{4\pi A\psi_1}{f_R}+\frac{4\pi\mu \dot{f}_R}{f_R^2}.
\end{align}
The geometric terms for the dark source in case of axial symmetry are
\begin{align}\nonumber
\mathcal{T}_{00}&=\frac{1}{\kappa}\left[\ddot{f}_R-\frac{\dot{f}_R^2 }{f_R}-\frac{A^2B'f_R'}{B^3}+\frac{f_R'C'}{B^2C}+\frac{1}{(B^2A^2r^2+G^2)}\left(A^2B^2r^2
\ddot{f}_R\right.\right.\\\nonumber&\left.\left.-B^2r^2A\dot{A}\dot{f}_R -\frac{A^3\dot{A}B^4r^4\dot{f}_R }{(B^2A^2r^2+G^2)}-G\dot{G}\dot{f}_R +\frac{A^2B^2r^2G\dot{G}\dot{f}_R}{(B^2A^2r^2+G^2)}-GAA_{,\theta}\dot{f}_R \right.\right.\\\nonumber&+\frac{A^2B^2r^2\dot{f}_R^2}{2f_R}\left.\left.-\frac{A^4B^4r^4\dot{f}_R^2 }{f_R(B^2A^2r^2+G^2)}-\frac{A^2B^2r^2G^2\dot{f}_R^2 }{2f_R(B^2A^2r^2+G^2)}-A^2B\dot{B}r^2\dot{f}_R\right.\right.\\\nonumber&\left.\left.-\frac{A^4B^3\dot{B}r^4\dot{f}_R}{ B^2A^2r^2+G^2}-\frac{A^2B\dot{B}r^2G^2\dot{f}_R}{B^2A^2r^2+G^2}-\frac{A^2GB_{,\theta}\dot{f}_R}{B}
+\frac{A^4GBB_{,\theta}r^2\dot{f}_R}{B^2A^2r^2+G^2}+\frac{A^4r^2f_R'^2}{2}\right.\right.\\\nonumber&\left.\left.
+\frac{A^2G^2f_R'^2}{2B^2f_R}-\frac{A^4r^2B'f_R'}{B}+A^3A'r^2f_R'+\frac{A^4B^2r^2G_{,\theta}\dot{f}_R}{B^2A^2r^2+G^2}
+A^4rf_R'+\frac{A^2GG'f_R'}{2B^2}\right.\right.\\\nonumber&\left.\left.-\frac{A^2B^2r^2\dot{C}\dot{f}_R}{C}
+\frac{GC_{,\theta}\dot{f}_R}{C}+\frac{3A^2B^2r^2\dot{f}_R^2}{4f_R}\right)\right]-\frac{A^2f_R}{2\kappa}\left(R-\frac{f}{f_R}\right)+\frac{3f_R'^2}{2\kappa B^2f_R}
\\\nonumber &-\frac{3\dot{f}_R^2}{2\kappa f_R},\\\nonumber
\mathcal{T}_{11}&=\frac{1}{\kappa}\left[-\frac{5f_R'^2}{4f_R}-\frac{C'f_R'}{C}+\frac{1}{B^2A^2r^2+G^2}\left(B^3\dot{B}r^2\dot{f}_R
+GBB_{,\theta}\dot{f}_R-\frac{B^4r^2\dot{f}_R^2}{2f_R}\right.\right.\\\nonumber&\left.\left.
+\frac{A^2B^6r^4\dot{f}_R^2}{f_R(B^2A^2r^2+G^2)}+\frac{B^4r^2G^2\dot{f}_R^2}{2f_R(B^2A^2r^2+G^2)}+
B^3\dot{B}r^2\dot{f}_R+\frac{A^2B^5\dot{B}r^4\dot{f}_R}{B^2A^2r^2+G^2}\right.\right.\\\nonumber&\left.\left.
+\frac{G^2B^3\dot{B}r^2\dot{f}_R}{B^2A^2r^2+G^2}
+G^3BB_{,\theta}-\frac{A^2B^2r^2f_R'^2}{2}-\frac{G^2B^2f_R'^2}{2B^2f_R}+A^2BB'r^2f_R'+B^4r^2\ddot{f}_R \right.\right.\\\nonumber&\left.\left.+\frac{A\dot{A}B^6r^4\dot{f}_R}{B^2A^2r^2+G^2}-\frac{B^4r^2G\dot{G}\dot{f}_R}
{B^2A^2r^2+G^2}-AA'B^2r^2f_R'-\frac{A^2B^4r^2G_{,\theta}\dot{f}_R}{B^2A^2r^2+G^2}-A^2B^2rf_R''\right.\right.\\\nonumber&\left.\left.
-\frac{GG'f_R'}{2}+\frac{B^4r^2\dot{C}\dot{f}_R}{C}-\frac{GB^2C_{,\theta}\dot{f}_R}{C}-\frac{3B^4r^2\dot{f}_R^2}{4f_R} \right)\right]-\frac{B^2f_R}{2\kappa}\left(R-\frac{f}{f_R}\right),\\\nonumber
\mathcal{T}_{12}&=\frac{1}{\kappa}\left[-\frac{B_{,\theta}f_R'}{B}+\frac{1}{B^2A^2r^2+G^2}\left(-\frac{B^2r^2G'\dot{f}_R}{2}
-GBB'r^2\dot{f}_R-GB^2r\dot{f}_R\right)\right],\\\nonumber
\mathcal{T}_{22}&=\frac{1}{\kappa}\left[rf_R'+\frac{r^2f_R'^2}{2f_R}-f_R''r^2-\frac{r^2C'f_R'}{C}+\frac{1}{B^2A^2r^2+G^2}\left(
B^2r^2G_{,\theta}\dot{f}_R\right.\right.\\\nonumber&\left.\left.+2GBB_{,\theta}r^2\dot{f}_R-\frac{5B^4r^4\dot{f}_R^2}{4f_R}
+\frac{A^2B^6r^6\dot{f}_R^2}{f_R(B^2A^2r^2+G^2)}+\frac{B^4r^4G^2\dot{f}_R^2}{2f_R(B^2A^2r^2+G^2)}\right.\right.\\\nonumber&\left.\left.
\frac{A^2B^5\dot{B}r^6\dot{f}_R}{B^2A^2r^2+G^2}+G^2B^3\dot{B}r^4\dot{f}_R-\frac{A^2B^3r^4GB_{,\theta}\dot{f}_R}{B^2A^2r^2+G^2}
-\frac{A^2B^2r^4f_R'^2}{2}-\frac{G^2r^2f_R'^2}{2f_R}\right.\right.\\\nonumber&\left.\left.+A^2BB'r^4f_R'+B^4r^4\ddot{f}_R
+\frac{A\dot{A}B^6r^6\dot{f}_R}{B^2A^2r^2+G^2}-\frac{B^4r^4G\dot{G}\dot{f}_R}{B^2A^2r^2+G^2}-AA'B^2r^4f_R'\right.\right.\\\nonumber &\left.\left.
-\frac{A^2B^4r^4G_{,\theta}\dot{f}_R}{B^2A^2r^2+G^2}-A^2B^2r^3f_R'-\frac{GG'r^2f_R'}{2}+\frac{B^4r^4\dot{C}\dot{f}_R}{C}
-\frac{GC_{,\theta}B^2r^2\dot{f}_R}{C}\right)\right]\\\nonumber &-\frac{B^2r^2f_R}{2\kappa}\left(R-\frac{f}{f_R}\right)+\frac{3f_r'^2}{2\kappa B^2f_R}.
\end{align}

\vspace{0.25cm}

\section*{Acknowledgments}

This work has been supported financially by National Research Project for Universities (NRPU), Higher Education Commission Pakistan under research Project No. 8769/Punjab/NRPU/R\&D/HEC/2017.

\end{document}